# Oxidation resistance of graphene-coated Cu and Cu/Ni alloy

*Shanshan Chen[‡,†,⊥], Lola Brown[§,⊥], Mark Levendorf[§], Weiwei Cai[†,‡], Sang-Yong Ju[§], Jonathan Edgeworth[‡], Xuesong Li[‡], Carl Magnuson[‡], Aruna Velamakanni[‡], Richard R. Piner[‡], Jiwoong Park[§,||,\*], Rodney S. Ruoff[‡,\*]*

[‡]Department of Mechanical Engineering and Texas Materials Institute, University of Texas at Austin, Austin, TX 78712, USA

[†]Department of Physics, Fujian Key Laboratory of Semiconductor Materials and Applications, Xiamen University, Xiamen 361005, China

[§]Department of Chemistry and Chemical Biology, Cornell University, Ithaca, NY 14853, USA

[||]Kavli Institute at Cornell for Nanoscale Science, Cornell University, Ithaca, NY 14853, USA

[⊥]These authors contributed equally to this work

\* Address correspondence to r.ruoff@mail.utexas.edu, jpark@cornell.edu.

## ABSTRACT

The ability to protect refined metals from reactive environments is vital to many industrial and academic applications. Current solutions, however, typically introduce several negative effects, including increased thickness and changes in the metal physical properties. In this paper, we demonstrate for the first time the ability of graphene films grown by chemical vapor deposition to protect the surface of the metallic growth substrates of Cu and Cu/Ni alloy from air oxidation. SEM, Raman spectroscopy, and XPS studies show that the metal surface is well protected from oxidation even after heating at 200 °C in air for up to 4 hours. Our work further shows that graphene provides effective resistance against hydrogen peroxide. This protection method offers significant advantages and can be used on any metal that catalyzes graphene growth.

The use of refined metals is widespread, but they are often chemically reactive, requiring protective coatings for many applications. Protecting the surface of reactive metals has developed into a significant industry which employs many different approaches, including coating with organic layers [1-3], paints or varnishes[4], polymers[5], formation of oxide layers[6], anodization[7], chemical modification[8], and coating with other metals or alloys[9]. However, these conventional approaches can suffer from a variety of limitations, such as susceptibility to damage by heat, limited chemical stability, cost, and formation of waste products. In addition, most conventional methods modify the physical properties of metals being protected. The addition of a protective coating changes the dimensions of the metal due to the finite thickness of the coating, changes the appearance and the optical properties of the metal surface, and often decreases the electrical and thermal conductivity. One important approach to overcome these problems would be to develop a novel protection coating with an exceptional chemical and thermal stability with minimum changes to the physical properties of the protected metal.

Graphene as a two-dimensional one-atom-thick sheet of carbon has attracted increased interest for both fundamental reasons and due to its potential for a wide variety of applications[10-13]. A world-wide effort is underway on developing new and improved methods of growing graphene on metal substrates; our group included[14-18]. In particular, chemical vapor deposition (CVD) techniques have been successfully applied to grow high quality single and multilayer graphene onto various metal substrates, including single-layer growth on Cu[17, 19], Pt[20], and Ir[21], and multilayer growth on Ni[15, 22] and Ru[23]. Growth temperature ranges from 650 °C to above 1000 °C, depending on the substrate of choice and the carbon source used[15, 17, 19-25]. While the CVD graphene films are primarily studied after they are removed from the growth substrates, we also examined the graphene-coated metal before dissolving the catalyst metal, and noticed that the metal had many steps and appeared very smooth[17], with surface topology significantly different from uncoated metal surfaces. Further investigation revealed that the differences in surface topology were due to the fact that graphene was protecting the metal substrate from oxidation[19].

The full potential of graphene as a protection layer can be understood based on its unique physical and chemical properties. First, surfaces of $sp^2$ carbon allotropes form a natural diffusion barrier thus providing a physical separation between the protected metal and reactants. This can be seen from the encapsulation of various atomic species inside of fullerenes [26] and carbon nanotubes at high temperatures and in vacuum[27]. More recently, graphene has been used to form a microscopic air-tight 'balloon'[28], which clearly demonstrates its property as an impermeable barrier. Second, graphene has exceptional thermal and chemical stability. Under an inert environment it is stable at extremely high temperatures (higher than 1500 ℃ [29-31]) and it is also stable under many conditions where other substrates would undergo rapid chemical reactions. In fact, the latter property has been the key to the processes used to separate large scale graphene from the substrates where they are grown. Combined, these two properties (impermeability and thermal/chemical stability) alone would make graphene an excellent candidate for a novel protection layer. Furthermore, graphene has several unique benefits. It is optically transparent (approximately 2.3% absorption per layer [32]) in visible wavelengths, electrically and thermally conductive, and it adds only about 0.34 nanometer per layer to the total dimension of the coated metal. To the best of our knowledge, the use of graphene as a passivation layer to protect metal surfaces has not been reported yet.

In this work, we demonstrate for the first time the ability of graphene films grown by CVD to protect the surface of the metallic growth substrate (Cu and Cu/Ni alloys) from oxidation, both in air at elevated temperatures, as well as in hydrogen peroxide. Large area graphene were grown directly on Cu foils, Cu/Ni alloys using methane as a carbon source by CVD. Details are discussed in the METHODS section. The performance of the graphene coating as a transparent, conductive and potentially passivating film on these metal foils was evaluated by heating the graphene-coated foils in an oven for 4 hours at 200 °C in laboratory air, as well as immersing 1 cm$^2$ pieces of such graphene-coated metal foils into a solution of 30% (weight/weight) hydrogen peroxide ($H_2O_2$, Fisher Scientific) up to 45 minutes. Two types of samples were prepared: graphene-coated Cu foil (designated as Cu+G), and graphene-coated Cu/Ni alloy foil (designated as Cu/Ni+G). These samples were compared to corresponding untreated metal foils (designated as Cu and Cu/Ni).

**RESULTS AND DISCUSSION**

Prior to any studies, micro-Raman spectroscopy was performed on each sample in order to verify the presence of high quality graphene films. Single layer graphene films display sharp G (~1580 cm$^{-1}$) and 2D (~2700 cm$^{-1}$) bands, with a low G/2D ratio (typically smaller than 0.5)[16, 17]. Multi-layer graphene can be identified by a high G/2D ratio (larger than or close to 1) and the altered shape of the 2D band[33, 34]. Defect density is indicated by the D (~1350 cm$^{-1}$)/G ratio[35], which is negligible for pristine single crystal graphene [36].

In Figure 1 we show photograph images of various metal surfaces, both graphene-coated and uncoated, after air anneals and exposure to liquid etchant. In all cases the graphene-coated metal surfaces show very little visible change, as opposed to the uncoated metals whose surfaces change appearance dramatically. As schematically shown in Figure 1a, the graphene film can be seen as a molecular diffusion barrier, preventing the reactive agent from ever reaching the metal underneath. More specifically, graphene-coated Cu and Cu/Ni foils show no changes after lengthy air anneals (200 °C, 4 hours, see Figure 1c), whereas uncoated films exhibited a substantial darkening. To further demonstrate the potential of graphene as a protection layer for bulk metal, we grew single layer graphene on a copper penny (95%Cu/5% Zn, minted 1962 - 1982). In Figure 1b two pennies are displayed, both of which were exposed to 30% $H_2O_2$ for 2 minutes. Although both pennies originally looked the same, a stark contrast arises between the graphene-coated (upper) and uncoated (lower) coins after exposure. The unprotected copper penny turned a dark shade of brown, whereas the protected coin maintained the original appearance. All these examples show that graphene passivates the growth surface, which, as discussed earlier, is due to its impermeability and chemical resistance. Below we discuss these two aspects in more detail.

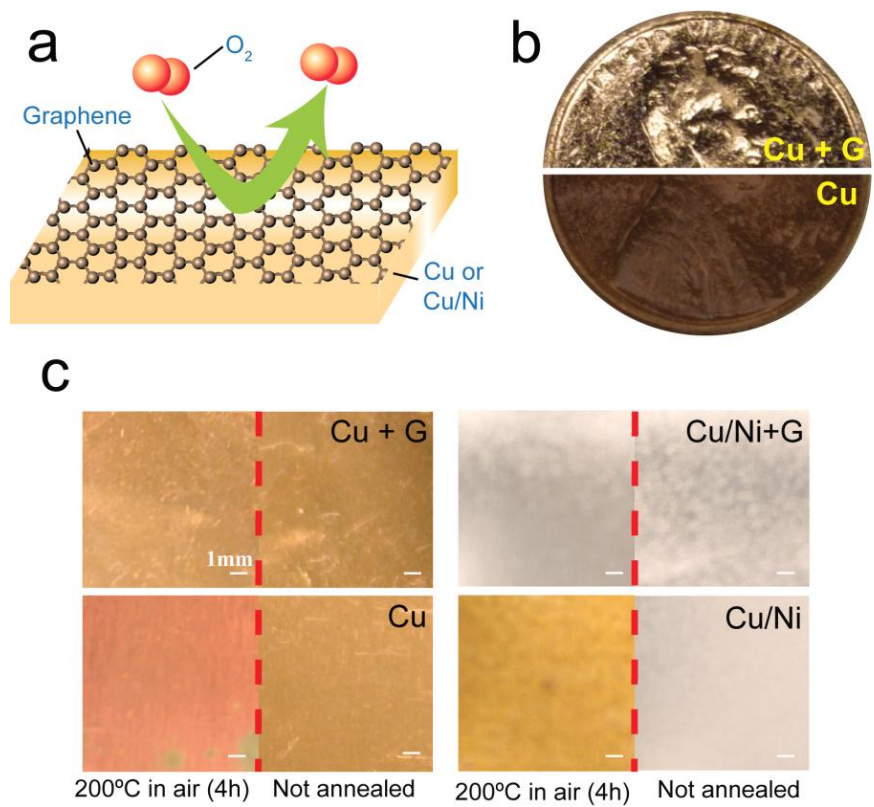

**Figure 1**. (a) Illustration depicting a graphene sheet as a chemically inert diffusion barrier. (b) Photograph showing graphene coated (upper) and uncoated (lower) penny after $H_2O_2$ treatment (30%, 2 min). (c) Photographs of Cu and Cu/Ni foil with and without graphene coating taken before and after annealing in air (200 °C, 4h).

Figure 2 shows SEM and XPS measurements of metal surfaces before and after air oxidation. The atomic steps under the graphene film are clearly visible for graphene coated samples before and after the anneals (Figure 2a and 2c, top), indicating that copper oxide has not formed beneath the graphene. The coated Cu and Cu/Ni surface is free from surface oxide due to the $H_{2(g)}$ exposure at a high temperature (1000 °C) prior to the growth of graphene. The metal suface is protected by the graphene layer during subsequent extended exposure at 200 °C in air. The micrograph has a number of small bright white spots representing oxides formed, most likely at the graphene grain boundaries or defect sites of the graphene surface, as we presented in a previous paper[18]. Better protection is afforded for the Cu/Ni alloy foil surface by a multilayer (as confirmed by the measured Raman ratio of the 2D/G band in Figure 3 for Cu/Ni+G (0h)) graphene coating. In contrast, images of unprotected metal foils after annealing show a rough surface structure and are much more blurry—likely due to a charging effect from the presence of oxides. It is difficult to obtain a clear image because of the accumulated charges in insulating oxides on the surface.

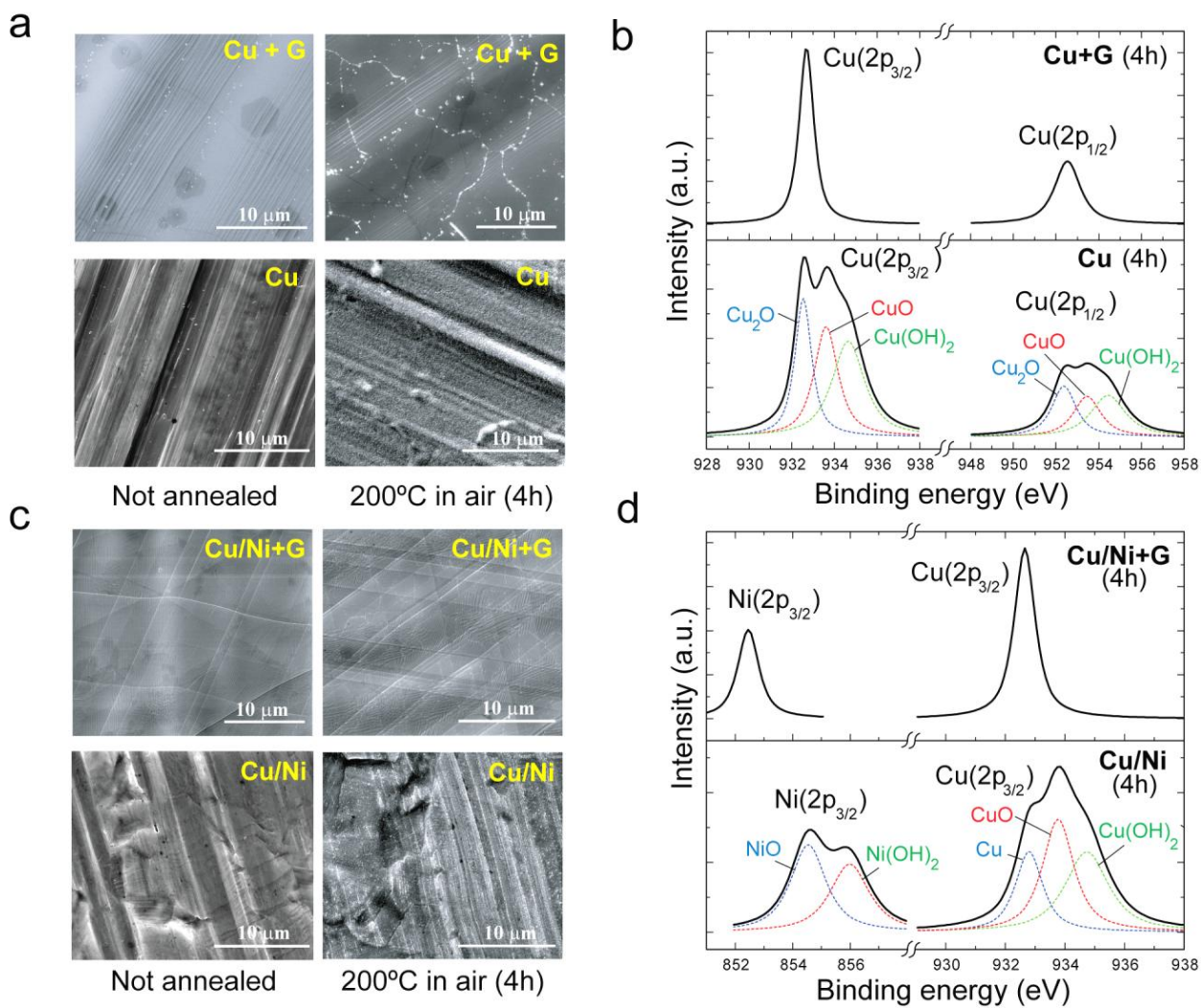

**Figure 2**. (a) SEM images of graphene coated (upper) and uncoated (lower) Cu foil taken before (left column) and after (right column) annealing in air. (b) XPS core-level Cu2p spectrum of coated (upper) and uncoated (lower) Cu foil after air anneal (200 °C, 4h). (c) SEM images of Graphene coated (upper) and uncoated (lower) Cu/Ni foil taken before (left column) and after (right column) annealing in air. (d) XPS core-level $Ni2p_{3/2}$ and $Cu2p_{3/2}$ spectrum of coated (upper) and uncoated (lower) Cu/Ni foil after air anneal (200 °C, 4h).

XPS was then performed on these substrates in order to provide an analysis of the metal composition after heat treatment (200 °C, 4h). The XPS spectrum of coated Cu foil before and after air annealing (200 °C, 4 h) both show two Cu peaks at binding energies of 932.6 and 952.5 eV, which correspond to $Cu2p_{3/2}$ and $Cu2p_{1/2}$ [37, 38] (Figure 2b). However, uncoated Cu foil shows broader peaks which correspond to different copper oxides, $Cu_2O$ (932.5 and 952.3 eV), CuO (933.6 and 953.4 eV), and $Cu(OH)_2$ (934.7 and 954.5 eV) [37, 38]. These data indicate that the graphene coating is clearly acting as a diffusion barrier, protecting the underlying copper from oxidation. Similarly, Figure 2d shows the XPS spectrum for the coated Cu/Ni foil. Two sharp peaks are present, corresponding to $Cu2p_{3/2}$ (932.6 eV) and $Ni2p_{3/2}$ (852.5 eV) [37], demonstrating no change in the chemical composition of the protected metal. As before, inspection of the uncoated foil reveals two broader peaks, one is comprised of two nickel oxide peaks, NiO (854.5 eV) and $Ni(OH)_2$ (856.0 eV) [37, 39], and the other is comprised of three peaks - metallic Cu (932.6 eV) and two copper oxide peaks, CuO (933.6 eV), and $Cu(OH)_2$ (934.7 eV). These XPS spectra demonstrate that the uncoated Cu/Ni foil was oxidized to a certain extent after heat treatment. The data in Figure 2d also provide a means to compare graphene as a protection layer with the Cu/Ni alloy inherent corrosion resistance. Upon oxidation the uncoated Cu/Ni alloy forms a protective film of $Cu_2O$ with Ni compounds (e.g., NiO) as minor components [40, 41]. This oxide layer is more stable due to the presence of Ni atoms in the copper lattice, resulting in a lower number of defects [40, 42]. The oxide therefore provides better protection against further oxidation, which explains the presence of a metallic Cu signal in Figure 2d (lower). Nevertheless, in our experiments the graphene-coated Cu/Ni alloy still shows significantly better oxidation resistance, compared to the uncoated Cu/Ni alloy, as can be seen from the absence of an oxide signals in Figure 2d (upper).

Under air oxidation graphene also shows remarkable chemical stability. Figure 3 presents the Raman spectra of coated and uncoated Cu and Cu/Ni foil samples, before and after heating in air (200 °C, 4 h). Before treatment, the coated Cu foil exhibits a large 2D/G peak ratio ~ 2 which is indicative of high quality single layer graphene[36]. The coated Cu/Ni foil also exhibits characteristics of high quality multilayer graphene – a low D band in conjunction with the distinct G and 2D peak shapes. After heat treatment, the uncoated Cu foil shows multiple peaks between 214 cm$^{-1}$ and 800 cm$^{-1}$, corresponding to various copper oxides - $Cu_2O$ (214, 644 cm$^{-1}$), CuO (299, 500 cm$^{-1}$) and Cu(OH)$_2$ (800 cm$^{-1}$) [43, 44]. Uncoated Cu/Ni foil displays CuO (299, 342, 634 cm$^{-1}$) and $Cu_2O$ (218 cm$^{-1}$) peaks, as well as NiO peaks (550 and 1100 cm$^{-1}$) [45, 46]. In contrast, the initial and final spectra of the coated foils are essentially identical. This clearly shows that the graphene is not only protecting the underlying metal, but is also virtually unaltered by the oxidizing gas. The data shown in Figures 2 and 3 illustrates that both single and multi-layer graphene serve as ideal protection coatings by both preventing diffusion and remaining chemically inert. Surprisingly, this oxidation protection by graphene is possible without strong adhesion to metal surfaces. Theory done to date indicates the interaction between graphene and the underneath metal is rather weak[47-50]. The graphene on Cu is considered to be physisorbed with a binding energy of $\Delta E < 0.07$ eV per carbon atom[47].

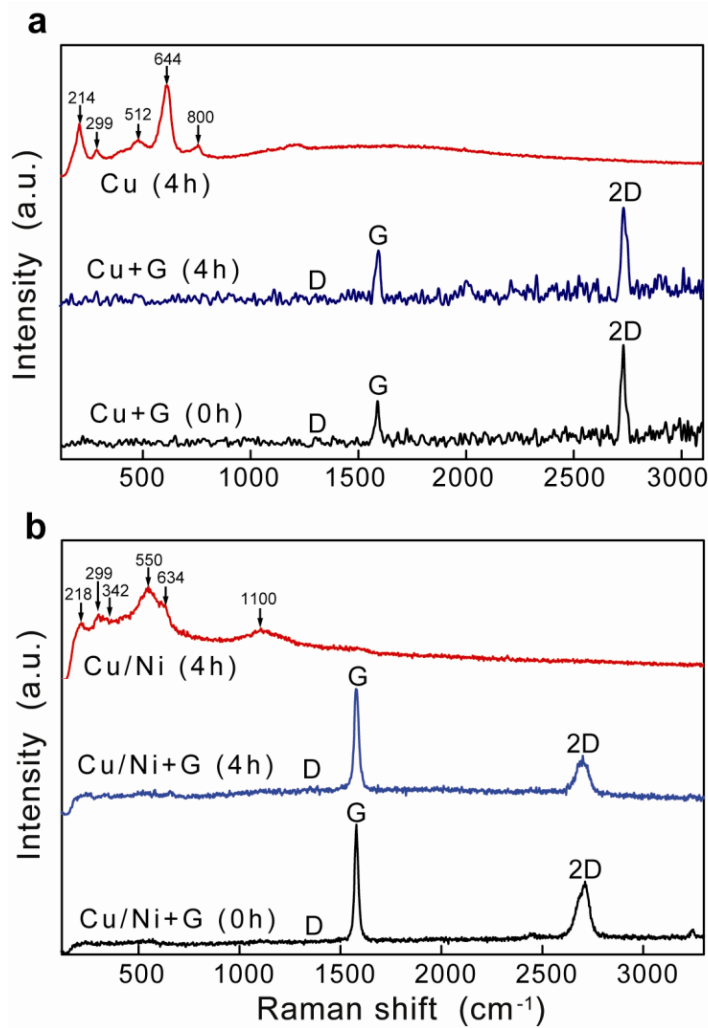

**Figure 3**. Raman spectrum of the Cu (a) and Cu/Ni alloy (b) foils with and without a graphene coating, acquired following heating in air at 200 ℃ for 0 and 4 hours, respectively.

To test longer term exposure, a run with 2 days exposure to air at 200 °C was performed on the graphene-coated Cu, and Cu/Ni alloy, samples. The Cu foils coated with monolayer graphene were oxidized to some extent, but a multilayer graphene-coated Cu/Ni foil remains visibly 'shiny'. The Raman spectrum obtained within graphene grains is identical to the 4 hour anneal, with sharp G and 2D Raman bands. However, regions of oxidized metal surface were formed along the grain boundaries of the graphene can seen by SEM, suggesting that the grain boundary is more susceptible to oxidative reactions.

Short time exposure to the oxidizing aqueous solutions $H_2O_2$ showed also significant protection for both graphene-coated Cu and Cu/Ni alloy foils. Graphene coated Cu and Cu/Ni samples were only attacked in few spots (white regions) after 15 and 5 minutes of $H_2O_2$ exposure respectively (Figure 4). Once the metal is exposed, the liquid can easily penetrate underneath the graphene sheet to attack the metal, since there is no oxide layer to slow the corrosion. Examples of this are seen in both the Cu and the Cu/Ni samples after longer $H_2O_2$ exposure, 45 and 15 minutes, respectively. The slower etch rate of graphene coated Cu than Cu/Ni might be due to the less reactive property of Cu than Ni in $H_2O_2$. The majority of the metal surface remains covered and protected by graphene.

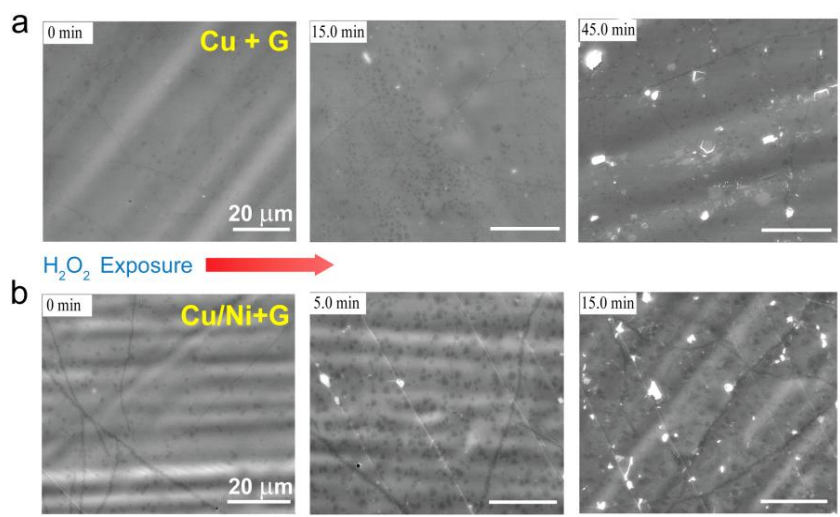

Figure 4. (a) SEM images of graphene coated Cu film after 30% $H_2O_2$ exposure for 0 min, 15 min and 45 min, respectively. (b) SEM images of graphene coated Cu/Ni alloy after 30% $H_2O_2$ exposure for 0 min, 5 min and 15 min, respectively.

In principle, "perfect" graphene – without defects and grain boundaries – is able to preserve the surface of metal under reactive environments over a long period of time thanks to its impermeability and chemical inertness. However, real CVD graphene is expected to show non-ideal behaviors. It is known that graphene is more likely to react at edges or where defects are present such as wrinkles, point defect as well as graphene boundaries[51]. Our results in Figure 2a suggest that grain boundaries are likely to be the main contributor to oxidation of the underlying metals. One way to solve this problem is to grow high quality and large domain single crystal graphene. Strong interest in nanoelectronics and other applications such as thermal management has already driven many groups around the world, our group included, to engineer grain structures and boundaries, and to rapidly increase graphene grain size. For example, we recently made progress on the growth of graphene with its grain size near one millimeter on Cu which will be reported soon[52]. Another way is to extend study to ultrathin graphite or hexagonal boron nitride thin films. The protection technique discussed here should work on any metals that can effectively catalyze graphene or ultrathin graphite growth, e.g., Cu, Ni, Fe, Ta, Pt, Ir, Ru and their alloys. Furthermore, refinement of graphene (multilayer graphene, ultrathin graphite) transfer techniques may allow their broader use on arbitrary substrates.

Further work is underway to explore graphene's functionality as an anti-corrosion protective coating for nano-electronic devices and other applications. This discovery would lead the world-wide graphene synthesis community to implement these novel applications of graphene, multilayer graphene, and ultrathin graphite as passivation coatings. The main limitation of this protection technique is its deactivation after mechanical damages. Therefore we would suggest using this protective coating on applications that do not involve circumstances where abrasion would ever be present, such as replacement of the Au coating for passivating Cu lines in semiconductor chip technology.

**CONCLUSIONS**

In this work, we demonstrate for the first time the excellent performance of graphene as a passivation layer. The ability of graphene coating to both prevent diffusion, as well as its chemical inertness to oxidizing gas and liquid solutions allow for its use in a wide variety of environments. Although partial oxidation may occur at graphene grain boundaries, we note that the graphene sheets provide near perfect protection within grains. With further advances in graphene growth and careful control of the metal catalyst, we anticipate a significant improvement in the level of protection these films may provide. Furthermore, refinement of graphene transfer techniques may even make it possible to take advantage of this material's amazing properties in any compatible system.

**METHODS**

**Preparation.** Large-area graphene samples were grown on metal substrates using previously reported CVD techniques [17, 19]. Growth of graphene on Cu foil (99.8% Alfa Aesar #13382) was performed in a hot wall tube furnace at a temperature of 1040 °C with 5% $H_2$ (ultra high purity; grade Air Gas, Inc.) in $CH_4$ (ultra high purity; grade Air Gas, Inc.) at a pressure of 500 mTorr. Growth of graphene on Cu/Ni (31% Ni, 67.8% Cu, 0.45% Mn, 0.60% Fe, else 0.15%, All Metal Sales, Inc.) was performed in the cold-wall CVD system at a growth temperature of 1000 °C with pure $CH_4$ at a pressure of 8 Torr.

**Characterization.** SEM images were taken with an FEI Quanta-600 FEG Environmental SEM using a voltage of 30 keV. XPS data was acquired to determine the chemical composition of the graphene films using a Kratos AXIS Ultra DLD XPS equipped with a 1808 hemispherical energy analyzer with photoemission stimulated by a monochromated Al Kα radiation source (1486.6 eV) at an operating power of 150 W. It was operated in the analyzer mode at 80 eV for survey scans and 20 eV for detailed scans of core level lines. Binding energies were referenced to the C 1s binding energy set at 284.5 eV. Curve fitting of each spectrum was performed using a Gaussian-Lorentzian peak shape after performing a Shirley background correction using Kratos Vision v2.2 software. Raman spectra (WITec Alpha 300) were obtained with a 532 nm laser (~50 mW power). AFM images were generated by a Park Scientific model CP "Research" (now, VEECO) with a contact force setting of 1 nanoNewton.

**Acknowledgements.** This work was supported by the Office of Naval Research and the DARPA Carbon Electronics for RF Applications Center, Center for Molecular Interfacing and Cornell Center for Materials Research – both funded by NSF, AFOSR PECASE. S.C. is supported by the China Scholarship Council Fellowship, and the National Natural Science Foundation of China. L.B. is supported by a Fulbright fellowship. M.L is supported by NSF IGERT training fellowship.